\documentclass[aps,prb,twocolumn,superscriptaddress,showpacs, longbibliography]{revtex4-1}
\usepackage{graphicx}
\usepackage{latexsym}
\usepackage{amsmath}
\usepackage{txfonts}
\usepackage{helvet}
\usepackage{braket}
\usepackage{amscd}

\usepackage{longtable}
\usepackage[%
  colorlinks=true,
  urlcolor=blue,
  linkcolor=blue,
  citecolor=blue
]{hyperref}

\begin{document}

\title{Occupation switching of $d$ orbitals probed via hyperfine interactions in vanadium dioxide} 

\author{Yasuhiro Shimizu}
\affiliation{Department of Physics, Graduate School of Science, Nagoya University, Furo-cho, Chikusa-ku, Nagoya 464-8602, Japan}
\author{Takaaki Jin-no}
\affiliation{Technical Center, Nagoya University, Furo-cho, Chikusa-ku, Nagoya 464-8601, Japan.}
\author{Fumitatsu Iwase}
\affiliation{Department of Physics, Tokyo Medical University, Shinjuku-ku, Tokyo 160-8402, Japan}
\author{Masayuki Itoh}
\affiliation{Department of Physics, Graduate School of Science, Nagoya University, Furo-cho, Chikusa-ku, Nagoya 464-8602, Japan}
\author{Yutaka Ueda}
\affiliation{Institute for Solid State Physics, University of Tokyo, Kashiwa, Chiba 277-8581, Japan}

\date{\today}

\begin{abstract} 
	Metal-insulator transition was microscopically investigated by orbital-resolved nuclear magnetic resonance (OR-NMR) spectroscopy in a single crystal of vanadium dioxide VO$_2$. Observations of the anisotropic $^{51}$V Knight shift and the nuclear quadrupole frequency allow us to evaluate orbital-dependent spin susceptibility and $d$ orbital occupations. The result is consistent with the degenerated $t_{2g}$ orbitals in a correlated metallic phase and the $d$ orbital ordering in a nonmagnetic insulating phase. The predominant orbital pointing along the chain facilitates a spin-singlet formation triggering metal-insulator transition. The asymmetry of magnetic and electric hyperfine tensors suggests the $d$ orbital reformation favored by a low-symmetry crystal field, forming a localized molecular orbital. The result highlights the cooperative electron correlation and electron-phonon coupling in Mott transition with orbital degrees of freedom. 
\end{abstract}

%\pacs{71.30.+h, 76.60.-k, 75.50.-y, 75.40.Gb}


\maketitle

\section{Introduction}
	Metal-insulator transition caused by an interplay among charge, spin, and orbital degrees of freedom has been one of the central issues in condensed matter physics \cite{Imada}. Transition metal compounds with orbital degrees of freedom exhibit the ground state determined by a delicate valance of electron correlations, Hund's exchange, intersite spin exchange, and spin-orbit couplings. The $d^n$ electron ($n$ =1--3, $t_{2g}$) systems under the octahedral crystal field are close to Mott-Hubbard insulator usually having antiferromagnetic ground states. An exceptional case is vanadium dioxide VO$_2$ exhibiting the metal-insulator transition into a nonmagnetic state through the unit-cell doubling \cite{Morin, Pouget, Kachi,Kawakubo}. The transition is driven via fertile ways such as applications of intense electric field, laser \cite{Cavalleri2004,Kubler,Liu}, pressure \cite{Baledent, Cheng2}, strain \cite{Mukherjee, Aetukuri, Park}, and doping \cite{Marezio, Booth,Chen,Nakano}, which are useful in practical applications to smart electronics. In contrast to the genuine Mott transition without symmetry breaking, the transition in VO$_2$ accompanies a structural distortion from a teragonal rutile ($R$, $P4_2/mnm$) metallic phase into a monoclinic ($M_1$, $P2_1/c$) insulating phase \cite{Andersson}. The antiferroelectric V-V pairing and twisting in the $M_1$ phase lower the symmetry of the ligand field and lift the orbital degeneracy of a $t_{2g}$ manifold \cite{Goodenough}. The structural change may involve a molecular orbital formation of the vanadium dimer \cite{Goodenough} or a $d$ orbital order \cite{Mott}. As spin-Peierls Mott insulator is adiabatically connected to band insulator, the driving force of metal-insulator transition in VO$_2$ has been debated for a long time, and extensive studies have concerned the fundamental issue. The system thus serves a crucial test for the precision of experiments and calculations. 

	\begin{figure}
	\includegraphics[scale=0.35]{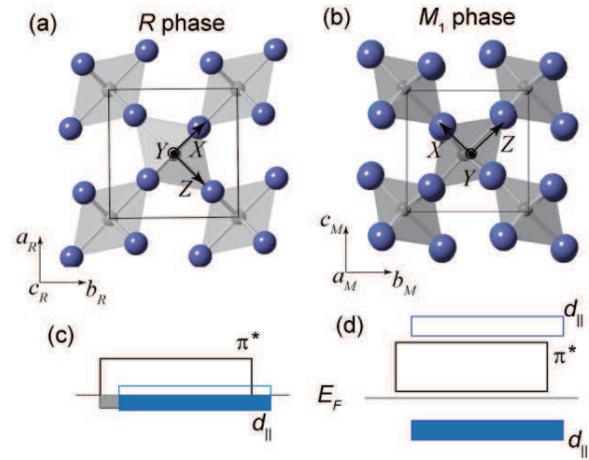}
	\caption{\label{Fig1} 
Crystal structure of VO$_2$ for (a) the tetragonal $R$ phase and (b) the monoclinic $M_1$ phase. Principal axes for the hyperfine coupling tensor point to $X = a_R + b_R$ $(c_M - b_M)$, $Y = c_R$ $(a_M),$ and $Z = b_R - a_R$ $(b_M + c_M)$ for the $R$ ($M_1$) phase. The tetragonal $P4_2/mnm$ structure contains two vanadium atoms connected by fourfold rotational symmetry in the unit cell. The edge-sharing VO$_6$ octahedra construct chains with the V-V distance of 2.85 ${\rm \AA }$ along the $c_R$ axis \cite{McWhan}. The dimerization in the monoclinic $P2_1/c$ structure results in alternating V-V distances, 2.65 and 3.12 ${\rm \AA}$, along the $a_M$ axis \cite{Andersson}. Schematic band structure for (c) the $R$ phase and (d) the $M_1$ phase. The coordinate $(X, Y, Z)$ is defined as the principal axis of the Knight shift and electric field gradient tensors. 
	}
	\end{figure}
Mott insulator is characterized by low-lying spin excitations, while opening large charge gap due to electron correlations. In the presence of dimerization, the spin excitation is also gapped at an energy scale much smaller than the charge gap. As originally pointed out by Mott, excitations in VO$_2$ are consistent with those expected in Mott insulator \cite{Mott} rather than band insulator having charge gap equivalent to spin gap \cite{Goodenough}. Indeed, the charge gap obtained from optical conductivity, $\Delta_{\rm c} = 0.66$ eV \cite{Shi, Okazaki,Eguchi}, and electrical conductivity, 0.6 eV \cite{Ladd}, is greater than the spin gap obtained from the $^{51}$V NMR spin-lattice relaxation rate, $\Delta_{\rm s} = 0.24$ eV \cite{Takanashi}. An emergence of the intermediate paramagnetic insulating phase (the $M_2$ or $T$ phase) under chemical doping \cite{Pouget} and uniaxial strains \cite{Gray} also supports the Mott-Hubbard picture \cite{Mott, Rice, Wentzcovitch, Huffman}. The band structure calculation based on the crystal structure of VO$_2$ shows a formation of antibonding and bonding orbitals in the $M_1$ phase, whereas the energy splitting is not large enough to open the band gap, implying a crucial role of electron correlation for opening the large charge gap. In the real space, the intradimer V-V distance 2.65 ${\rm \AA}$ in the $M_1$ phase is shorter than the metallic phase, 2.85 ${\rm \AA}$. In contrast, the interdimer separation 3.12 ${\rm \AA}$ is much longer and exceeds a critical value ($R_c \sim 2.9$ ${\rm \AA}$) to form a bonding band \cite{Goodenough1960}. It suggests the coexistence of Peierls and Mott regimes in the electron cloud distribution.

The orbital degree of freedom plays a key role in the metal-insulator transition of VO$_2$, as initially marked by Goodenough \cite{Goodenough}. Under the VO$_6$ distortion in the tetragonal $R$ phase, a $t_{2g}$ triplet of V$^{4+}$ ($d^1$) is lifted into $d_{||}$ (also termed as $d_{xy}$ or $\sigma$ orbital) directed along the chain, and $\pi^*$ ($d_{yz}$ and $d_{zx}$ orbitals) hybridized to oxygen $p$ orbitals \cite{Goodenough, Eyert}, as shown in Fig. \ref{Fig1}. The $d_{||}$ band overlaps the $\pi^*$ band in the $R$ phase [Fig. \ref{Fig1}(c)] and becomes lower in the $M_1$ phase by forming a bonding orbital band [Fig. \ref{Fig1}(d)]. The electron correlation may also split the narrower $d_{||}$ band into lower and upper Hubbard bands. Thus the lower-lying orbital has conflicting characters of the bonding orbital and the lower Hubbard band. Theoretical calculations with a local density approximation (LDA) and a dynamical mean-field theory (DMFT) account for the metal-insulator transition by including strong electron correlations \cite{Eyert, Liebsch, Biermann, Brito,Gatti,Kim}. Another theory is the orbitally-driven or correlation assisted Peierls transition \cite{Liebsch, Biermann}, supporting the $d$ occupation switching expected in optical, x-ray absorption, and diffraction measurements \cite{Haverkort, Okazaki, Qazilbash,Cavalleri,Budai}. Indeed, the $d$ orbital is mostly occupied in $d_{||}$ \cite{Haverkort}, consistent with a quasi one-dimensional (1D) structure with the Peierls instability. In contrast, a recent theoretical calculation shows hybridization to oxygen $p$ orbitals to form a molecular orbital in the $M_1$ phase \cite{Zheng}, constructing three-dimensional transfer paths. An accurate determination of the electron cloud distribution is required for determining the electronic structure strongly dependent on the orbital occupation. 
	
	Nuclear magnetic resonance (NMR) is a local probe to detect symmetry breaking and provides microscopic information about magnetic, orbital, and multipole orders \cite{Abragam, Kiyama, Takigawa, Tokunaga}. In transition-metal atoms, electric quadrupole and dipole hyperfine interactions are governed by anisotropic electron and spin density distributions, respectively \cite{Abragam, Kiyama}. In metallic vanadium oxides having one or two electrons in $d$ orbitals, the anisotropic hyperfine interaction represents $d$ orbital occupations \cite{Shimizu, Shimizu2, Shimizu3, Shimizu5}. As for VO$_2$, previous NMR measurements were conducted only on the powder sample and hence unable to quantitatively evaluate the orbital occupation \cite{Pouget, Takanashi}. 
	
In this paper, we report orbital-resolved $^{51}$V NMR (OR-NMR) spectroscopy on a single crystal of VO$_2$. Through $^{51}$V Knight shift and nuclear quadrupole frequency measurements, we extract $d$ orbital contributions in magnetic and electric hyperfine coupling tensors. The result uncovers the $d$ orbital shape in each phase. In the following, we will describe our experimental and analysis methods for OR-NMR in the section II, and experimental results for the metallic and insulating phases in the section III, followed by the discussion based on theoretical calculations in the section IV.

\section{Methods}
\subsection{Experimental setup}
	Single crystals of VO$_2$ were grown by a chemical transport method \cite{Kachi}. The typical dimension of the crystal was 2 mm $\times$ 2 mm $\times$ 5 mm. The crystal axes were determined from the x-ray diffraction pattern and the angular profile of the $^{51}$V NMR spectrum. Magnetic susceptibility was measured with a magnetometer (MPMS-XS, Quantum Design Ltd.) at 1 T. We obtained frequency-swept $^{51}$V NMR spin-echo spectra using $\pi/2$ pulses (1 $\mu$s length) with an interval time $\tau = 10 - 20$ $\mu$s at a constant magnetic field $H_0$ = 5.8701 T. The angular dependence of the $^{51}$V NMR spectrum was measured with a dual axis goniometer in the metallic $R$ phase at 350 K and in the insulating $M_1$ states at 300 K. 
	
\subsection{OR-NMR}
	For $d^1$ systems with weak spin-orbit coupling, the magnetic hyperfine interaction $\mathcal {H}_{\rm mag}$ is given by
	\begin{eqnarray}
	\mathcal {H}_{\rm mag} = -\mathcal {P}[\kappa {\bf S\cdot I} + \xi {\bf S} \cdot {\bf {\sf q}} \cdot {\bf I}] - \gamma_N \hbar \bf H_0 \cdot \alpha \cdot \bf I, 
	\label{Eq1}
	\end{eqnarray} 
	where $\mathcal {P}$ represents the product of Bohr magneton $\mu_{\rm B}$, nuclear gyromagnetic ratio $\gamma_{\rm N}$, and an ionic radial average factor $\braket {r^{-3}} = 3.684$ a.u. for a free V$^{4+}$ ion ($3d^1$) \cite{Abragam}. Here $\hbar$ is Plank's constant, $\kappa$ a coefficient of Fermi contact interaction ($\simeq 0.5$ for vanadates) \cite{Abragam}, ${\bf S}$ $({\bf I})$ electron (nuclear) spin operator, and ${\bf H_0}$ external magnetic field. The first term in Eq.(\ref{Eq1}) comes from core polarization of inner $s$ electrons, giving isotropic Knight shift proportional to spin susceptibility $\chi^{\rm s}$: $K^{\rm s}_{\rm iso} = A^{\rm s}_{\rm iso}\chi^{\rm s}/N\mu_B$ with the Avogadro's number $N$ and the isotropic hyperfine coupling constant $A^{\rm s}_{\rm iso} = -\kappa \mu_{\rm B}\braket {r^{-3}}$. The second term in Eq.(\ref{Eq1}) is the equivalent operator expression of dipole hyperfine interaction, where $\xi = \frac{2}{21}$ for $d^1$, and ${\sf q}$ denotes the quadratic tensor having components $q_{ij} \equiv \frac{3}{2}(L_i L_j + L_j L_i ) - \delta_{ij}{\bf L}^2$ ($i, j$ = $x, y, z$), where ${\bf L}$ is the total orbital angular momentum \cite{Abragam}. The orbital quadrupole moment tensor ${\sf q}$ is simply expressed in terms of the orbital equivalent operator and reflects the $3d$ orbital occupation. For example, the diagonal components of ${\sf q}$ for $d_{xy}$ are given by $q_{zz} = -2q_{xx} = -2q_{yy} = 6$ \cite{Autschbach}, where the coordinate ($x, y, z$) for $d$ orbitals is taken along the V-O bond direction, which is rotated by 45$^\circ$ from the principal axis ($X, Y, Z$) along the $Z$ axis. The even occupation of $t_{2g}$ vanishes the ${\sf q}$ components. For the arbitrary occupation ratio $d_{xy} : d_{yz} : d_{zx} = a: b: c$ $(a + b + c = 1)$, the diagonal components of the dipolar Knight shift tensor, divided by $K^{\rm s}_{\rm iso} = (K_X^s + K_Y^s + K_Z^s)/3$, is expressed as $\frac{4}{7}(-a-b+2c, -a+2b-c, 2a-b-c)$ \cite{Shimizu2, Takeda}. Thus the axial Knight shift $K_{\rm ax}^s = (2K_Z^s - K_X^s - K_Y^s)/3$ and the asymmetric shift $K_{\rm as}^s = (K_X^s - K_Y^s)/2$ divided by $K^{\rm s}_{\rm iso}$ are given by $\frac{4}{7}(2a-b-c)$ and $\frac{6}{7}(-b+c)$, respectively. The anisotropy of the hyperfine coupling tensor can also give the occupation ratio using the similar relation to the Knight shift.  
	
	The third term in Eq. (\ref{Eq1}) represents the Van-Vleck process arising from the second order Zeeman effect due to spin-orbit coupling, which leads to temperature-independent orbital susceptibility. The components of ${\bf \alpha}$ are expressed as $\alpha_{\mu \nu} = 4\mu_{\rm B}^2\braket {r^{-3}}\Lambda_{\mu \nu}$, where $\Lambda_{\mu \nu}$ ($\mu$, $\nu$ = $x$, $y$, $z$) is the mixing element of $L_\mu L_\nu$ between the ground and excited states. 

For the nuclear spin $I = 7/2$, the NMR spectrum is split into seven through the electric quadrupole interaction $\mathcal {H}_{\rm el}$ between $^{51}$V nucleus quadrupole moment ($Q$ = $-0.05 \times 10^{-24}$ cm$^2$) and electric field gradient (EFG) ${\sf V}$. The electric hyperfine interaction with anisotropic $d$ orbitals is expressed as
	\begin{eqnarray}
	\mathcal {H}_{\rm el} = \frac{e^2Q}{2I(2I-1)}\xi \braket {r^{-3}}[{\bf I}\cdot {\sf q} \cdot {\bf I}],
	\label{Eq2}
	\end{eqnarray} 
	where the quadratic tensor ${\sf q}$ is equal to that appeared in the dipole hyperfine coupling. The nuclear quadrupole splitting frequency tensor $\delta \nu$ is written by using the EFG tensor:  
	\begin{eqnarray}
	{\sf \delta \nu} = \frac{3eQ}{2hI(2I-1)}{\sf V}. 
	\label{Eq3}
	\end{eqnarray} 
	The onsite contribution to the quadrupole frequency due to anisotropic $d$ electrons is given by
	\begin{eqnarray}
	\delta \nu^d = \frac{e^2Q}{7hI(2I-1)}\braket {r^{-3}}{\sf q}.
	\label{Eq4}
	\end{eqnarray} 
Here the full $d$ orbital polarization gives the maximum principal component $\delta \nu^d_Z$ at $|q_{ZZ}| = 6$. Using the $d$ occupation ratio for three $t_{2g}$ orbitals, $\delta \nu^d$ is expressed as $-0.71(-a-b+2c, -a+2b-c, 2a-b-c)$ MHz. Therefore, we can obtain the $d$ occupation from the $\delta \nu$ tensor by subtracting the outer ion contribution. 

\subsection{Density functional calculations}
The EFG tensor ${\sf V}_{\rm cal}$ at $^{51}$V nucleus was calculated by the linearized augmented plane wave + local orbital (LAPW + lo) method based on the density functional theory (DFT) implemented in the WIEN2k code \cite{Blaha}. The exchange correlation functional is Perdew-Burke-Ernzerhof
(PBE) derived from the generalized gradient approximation (GGA) \cite{Perdew}. We considered the relaxation of all atomic positions with the optimization of the unit cell volume. The on-site Coulomb interaction $U$ was considered for the insulating phase with the typical effective Coulomb interaction $U_{\rm eff} = U - 2J$ = 2.6 eV, $U = 4.2$ eV, and the Hund exchange coupling $J = 0.8$ eV \cite{Biermann, Liebsch}. The calculated results reproduce the previous reports \cite{Laad, Biermann, Zheng}. We also found that the band gap opens without including $U$ by utilizing the structure optimization.

\section{Experimental results}
\subsection{Metal-insulator transition}
	\begin{figure}
	\includegraphics[scale=0.5]{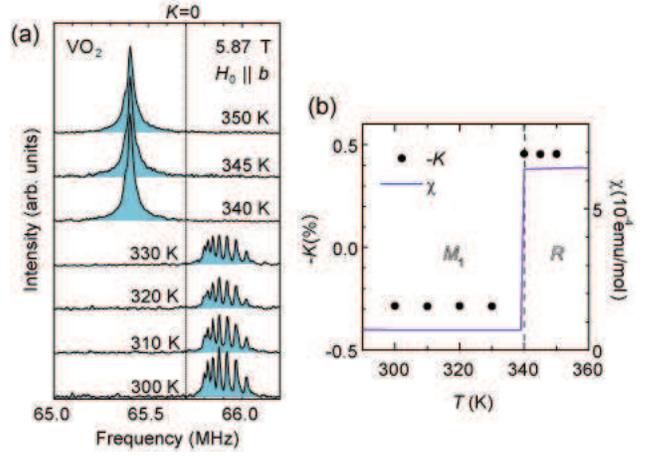}
	\caption{\label{Fig2} 
(a) $^{51}$V NMR spectrum for a constant field $H_0 (5.87$ T) $\parallel b$ across the metal-insulator transition at 340 K in VO$_2$. A vertical dotted line at 65.704 MHz represents the Knight shift origin $K=0$ ($^{51}\gamma = 11.193$ MHz/T). (b) Temperature dependence of $^{51}$V Knight shift $K$ (solid circles, the left-hand axis) and magnetic susceptibility $\chi$ (solid line, the right-hand axis), where the core diamagnetic susceptibility $\chi^{\rm dia} = -7 \times 10^{-6}$ emu mol$^{-1}$ is subtracted from the bulk susceptibility. 
	}
	\end{figure}
	Figure \ref{Fig2}(a) shows the temperature dependence of the $^{51}$V NMR spectrum measured for a constant magnetic field $H_0$ along the $b_R$ ($b_M$) axis in the $R$ ($M_1$) phase. We find a sharp single line in the $R$ phase above 340 K for the field direction close to a magic angle where the nuclear quadrupole splitting vanishes. A negative frequency shift ($K = -0.45\%$) in the $R$ phase despite positive $\chi$ indicates a negative hyperfine coupling constant due to the predominant core polarization in Eq.(\ref{Eq1}). The Knight shift behaves nearly invariant against temperature, consistent with Pauli paramagnetic susceptibility ($\chi = 6.4 \times 10^{-4}$ emu mol$^{-1}$) in the bulk measurement. From the $K$-$\chi$ plot, a linear scaling factor of the isotropic hyperfine coupling constant $A_{\rm iso}$ = $-8.3(4)$ T/$\mu_{\rm B}$, in agreement with that obtained for powder samples \cite{Pouget,Takanashi}, where the number inside the parentheses denotes the experimental uncertainty.  

Below 340 K, the NMR spectrum shows a dramatic change in the Knight shift and the quadrupole splitting. The Knight shift changes into a positive value, indicating the vanishing spin susceptibility $\chi^{\rm s}$ due to spin-singlet formation. A residual temperature-independent shift is attributable to the Van-Vleck orbital susceptibility, $K^{\rm orb}$ = 0.40(3)\%, corresponding to $\chi^{\rm orb} = 1.3(2) \times 10^{-4}$ emu mol$^{-1}$ using the orbital hyperfine coupling $A^{\rm orb} = 2\mathcal {P}$ = $46$ T/$\mu_{\rm B}$ for the free V$^{4+}$ ion \cite{Ohama, Abragam}. The finite $\chi^{\rm orb} > 0$ supports the Mott insulating picture, whereas $\chi^{\rm orb}$ should be diamagnetic in band insulators. An emergence of large quadrupole splitting reflects orbital ordering and a slight tilting of the principal axis by $\sim 2^\circ$ due to the drastic lattice distortion in the $M_1$ phase. The splitting is not equally spaced because of the small second order effect of the nuclear quadrupole interaction appearing apparently at the proximity of the magic angle.

\subsection{$R$ phase}
	\begin{figure}
	\includegraphics[scale=0.68]{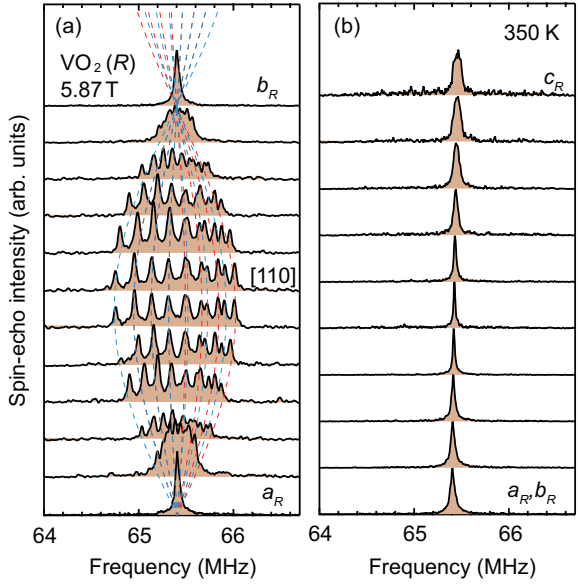}
	\caption{\label{Fig3} 
	(a) Angular dependence of $^{51}$V NMR spectrum in a single crystal of VO$_2$ for the $R$ phase at 350 K. The spectral intensity was normalized at the peak position for each angle. The magnetic field was rotated from $a_R$ to $b_R$ with a $9^\circ$ step. Red and blue curves trace resonance lines for two vanadium sites satisfying fourfold rotational symmetry. The central line gives the Knight shift $K$, while $\delta \nu$ is defined by the averaged splitting frequency. (b) The angular dependence of the spectrum from $a_R$ or $b_R$ to $c_R$ axis taken by a $10^\circ$ step. 
	}
	\end{figure}
	Figure \ref{Fig3} displays the angular dependence of the $^{51}$V NMR spectrum in the $R$ phase. Two vanadium atoms in the tetragonal unit cell are related by fourfold rotational symmetry with glide planes normal to $a_R$ and $b_R$ axes, and thus identical at the field direction along the glide planes, otherwise the number of resonance lines doubles. Indeed, we observed two sets of the quadrupolar split (14 lines at a maximum) for $I = 7/2$, as the magnetic field is rotated from the $a_R$ to $b_R$ axis. The principal axes $(X, Y, Z)$ of the Knight shift and EFG tensors respectively point to [110], [001], and [${\bar 1}$10] for a vanadium site located at the center of the unit cell, as shown in Fig. 1(a). The principal $Z$ axis is defined by the direction that gives a maximum of $|K|$ and $|\delta \nu|$. It corresponds to one of V-O bond directions, as expected for the small tetragonal VO$_6$ distortion along [${\bar 1}$10]. 
	
	\begin{figure}
	\includegraphics[scale=0.48]{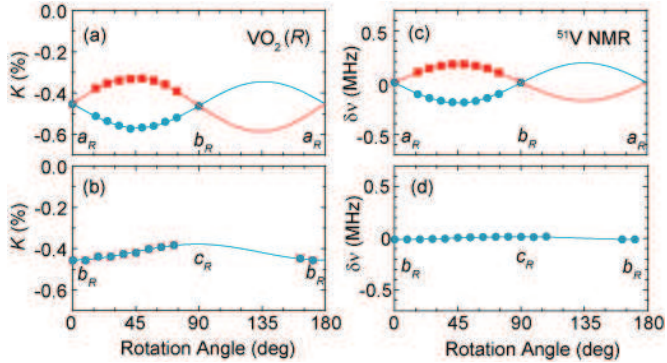}
	\caption{\label{Fig4} 
	Angular dependence of (a,b) $^{51}$V Knight shift $K$ and (c,d) quadrupole splitting frequency $\delta \nu$ in the $R$ phase of VO$_2$ at 350 K. The dotted curves are fitting results with Eq. (\ref{Eq4}). 
	}
	\end{figure}
	For the $ac$ plane rotation, the $^{51}$V NMR spectrum exhibits weak angular dependence without splitting [Fig. \ref{Fig3}(b)]. The identical angular dependence is expected in the $bc$ plane rotation for the tetragonal lattice. Thus we obtained the angular dependence of $K$ and $\delta \nu$, as displayed in Fig. \ref{Fig4}. To obtain the tensor, the angular dependence of $K$ is analyzed with sinusoidal functions with fitting parameters $k_1$, $k_2$, and $k_3$ \cite{Slichter},  
	\begin{eqnarray}
	K = k_1 {\rm cos}2\theta + k_2 {\rm sin}2\theta + k_3. 
	\label{Eq5}
	\end{eqnarray} 
	The Knight shift tensor ${\sf K}$ is determined as 
\begin{equation}
{\sf K} =
%\begin{align}
	\begin{pmatrix}
	    -0.34&0&0\\0&-0.38&0\\0&0&-0.58\\ 
	\end{pmatrix}
	\% .
%\end{align}
\label{Eq6}
\end{equation}

The result gives the isotropic shift $K_{\rm iso} = -0.43(2)\%$. Here the Knight shift consists of the spin and orbital components. The orbital part governed by the Van-Vleck orbital susceptibility can be evaluated as ${\sf K}^{\rm orb} = (0.64(5), 0.47(4), 0.20(3))\%$ based on the $K-\chi$ plot as an implicit function of temperature. After subtracting the orbital term, we obtained the spin part of the Knight shift tensor as ${\sf K}^{\rm s} = (-0.98(5), -0.85(4), -0.78(5))\%$. 
	
To extract the anisotropic hyperfine interaction, we evaluated the isotropic shift $K_{\rm iso}^s = -0.87(4)\%$, the axial anisotropy $K_{\rm ax}^s = 0.09(3)\%$, and the asymmetry $K_{\rm as}^s = -0.06(3)\%$. We find a sizable reduction of $K_{\rm ax}^s$ and $K_{\rm as}^s$ compared with $K_{\rm iso}^s$. It represents the nearly isotropic $d$ orbital shape due to the even occupation of $t_{2g}$ orbitals in the $R$ phase. As $K_{\rm iso}^s$, $K_{\rm ax}^s$, and $K_{\rm as}^s$ are all proportional to $\chi^{\rm s}$, the ratios $K_{\rm ax}^s/K_{\rm iso}^s = -0.10(3)$ and $K_{\rm as}^s/K_{\rm iso}^s = -0.07(2)$ cancel $\chi^{\rm s}$ and scale to the effective anisotropy of ${\sf q}$ expressed as a linear combination of ${\sf q}$ for three $t_{2g}$ orbitals. Taking the occupation as a parameter, the ratio is evaluated as $d_{||}: d_{yz}: d_{zx} = 0.27(4) : 0.41(5): 0.32(4)$. 

The magnetic hyperfine anisotropy measures the orbital-dependent spin polarization that scales to the partial density of states at the Fermi level for metallic systems. The $ab$ $initio$ band structure shows the total density of states of 1.8 state/eV, ($d_{||}: d_{yz}: d_{zx} = 0.845 : 0.649 : 0.340$) giving Pauli paramagnetic susceptibility of $1.2 \times 10^{-4}$ emu mol$^{-1}$ for free electrons. The difference to the experimental result may come from the orbital dependent spin correlation. In comparison with the other metallic vanadates, $K_{\rm ax}^s/K_{\rm iso}^s$ is much smaller than those of V$_6$O$_{13}$ (0.4--0.7) \cite{Shimizu} and LiV$_2$O$_4$ (0.7) \cite{Shimizu2} showing orbital dependent localization but close to the hollandite vanadate K$_2$V$_8$O$_{16}$ (0.1) having the edge-sharing VO$_6$ chain similar to VO$_2$ \cite{Shimizu6}. Thus the orbital degeneracy can be a manifestation of the weakly correlated metal in transition metal compounds.   

	\begin{figure}
	\includegraphics[scale=0.48]{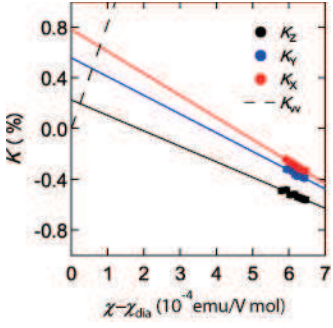}
	\caption{\label{Fig5} 
	Knight shift $K$ plotted against the bulk susceptibility $\chi$ for the principal axes of the Knight shift tensor \cite{Pouget}. The dashed line represents the calculated orbital shift for the free ion \cite{Abragam}. 
	}
	\end{figure}
	The occupation obtained from the Knight shift anisotropy postulates the isotropic spin susceptibility, which is valid for a system with the isotropic $g$ value. To confirm the result, we directly evaluated the anisotropy of the hyperfine coupling tensor ${\sf A}$ from the $K$-$\chi$ plots as an implicit function of temperature for the principal axes, as shown in Fig. \ref{Fig5}, in reference to the result of the powder sample \cite{Pouget}. We obtained the diagonal components of ${\sf A} = (-9.7(8), -8.3(7), -6.8(5))$ T/$\mu_{\rm B}$, yielding the isotropic part of the hyperfine coupling $A_{\rm iso} = -8.3(7)$ T/$\mu_{\rm B}$, the axial part $A_{\rm ax}/A_{\rm iso} = -0.17(2)$, and the asymmetric part $A_{\rm as}/A_{\rm iso} = 0.09(1)$. The small anisotropy is also consistent with the even contribution of $t_{2g}$ orbitals to the spin susceptibility. The population ratio is evaluated as $d_{||}: d_{yz}: d_{zx} = 0.23(3):0.44(5):0.33(4)$, in good agreement with the result obtained from the Knight shift anisotropy, which manifests the $d$ orbital degeneracy due to the small tetragonal ligand field. 

Another measure of the $d$ orbital occupation is the nuclear quadrupole splitting $\delta \nu$. Here we defined $\delta \nu$ as an average of six split interval frequencies in the spectrum. The angular dependence of $\delta \nu$ was analyzed with the sinusoidal function with fitting parameters $\nu_1$, $\nu_2$, and $\nu_3$ \cite{Volkoff}, 
	\begin{eqnarray}
	\delta \nu = \nu_1 {\rm cos}2\theta + \nu_2 {\rm sin}2\theta + \nu_3. 
	\label{Eq7}
	\end{eqnarray} 
As shown in Fig. 4(c, d), Eq.(\ref{Eq7}) well fits the experimental result. After diagonalizing the tensor, we obtained $\delta \nu = (187(6), 16(5), -203(7))$ kHz, where the principal axes are identical to those of ${\sf K}$. The asymmetry parameter is obtained as $\eta = |\delta \nu_X - \delta \nu_Y|/\delta \nu_Z$ = 0.84(2). 

	\begin{figure*}
	\includegraphics[scale=0.85]{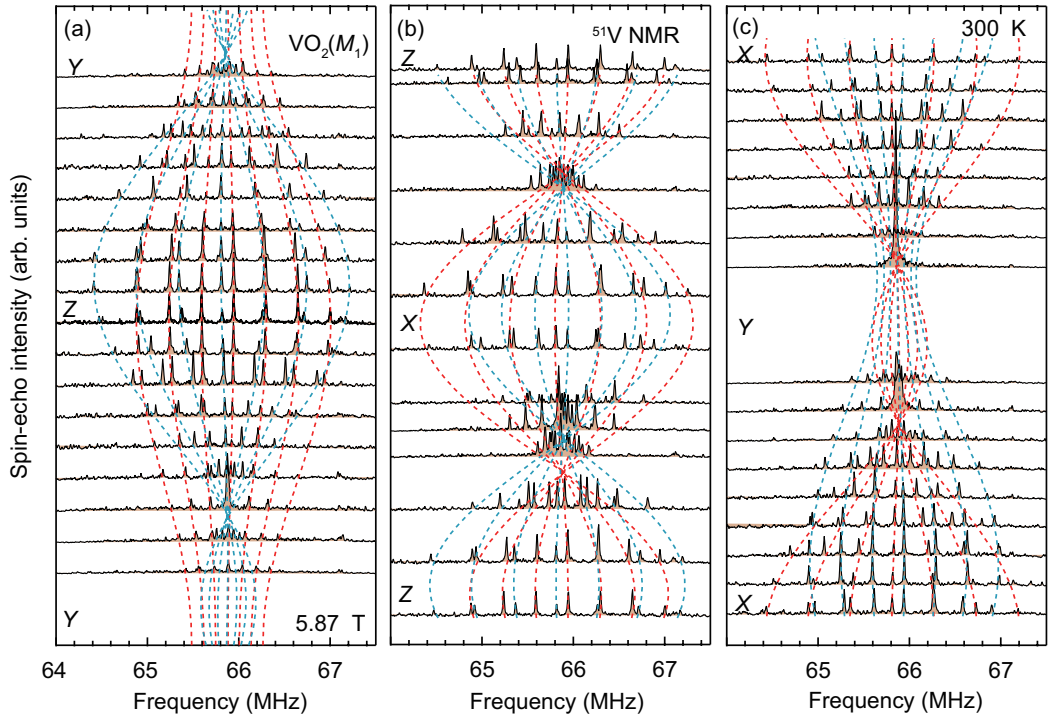}
	\caption{\label{Fig6} 
	Angular dependence of $^{51}$V NMR spectrum for the insulating $M_1$ phase at 300 K in VO$_2$. The spectrum was taken by a $9^\circ$ step with a two-axis goniometer. The dotted curves are fitting results with the sinusoidal function. The principal axes ($X, Y, Z$) for blue curves correspond to ($b_M + c_M$, $a_M$, $c_M - b_M$) in the crystal structure of the $M_1$ phase. 
	}
	\end{figure*}

In general $\delta \nu$ comes from a sum of EFG due to on-site $d$ orbitals, ${\sf V}^d$, and that of surrounding ions, ${\sf V}^{\rm ion}$. The latter contribution to $\delta \nu$ can be evaluated as $\delta \nu^{\rm ion} = (4, -18, 14)$ kHz with a point-charge calculation based on the crystal structure at 350 K \cite{McWhan}. The result roughly scales to the first-principles DFT calculation, ${\sf V} = (0.36, 2.11, -2.47)\times 10^{21}$ V/m$^2$, corresponding to $\delta \nu^{\rm cal} = (-31, -182, 213)$ kHz. By subtracting $\delta \nu^{\rm ion}$ from $\delta \nu$ using a anti-shielding factor 10 \cite{Sternheimer, Abragam}, the $d$ orbital contribution to the quadruple frequency is obtained as $\delta \nu^d = (-227, 164, 63)$ kHz. In a similar manner to the Knight shift, the orbital occupation ratio is evaluated as $d_{||}: d_{yz}: d_{zx} = 0.49(4): 0.24(3) : 0.26(3)$, which are insensitive to the anti-shielding factor within the experimental uncertainty. Thus the EFG anisotropy measuring the net orbital occupation is also consistent with the degenerated $t_{2g}$ orbitals. Whereas the lattice symmetry is tetragonal, the local orthorhombic VO$_6$ distortion can lift the degeneracy of the $\pi^*$ band and lead to the different occupation in $d_{yz}$ and $d_{zx}$. 
	
\subsection{Insulating $M_1$ phase}
	In the monoclinic $M_1$ phase, the unit cell contains four vanadium ions equivalent under magnetic field along crystal axes. Two of them become inequivalent for the field normal to the mirror plane. Two vanadium sites forming a dimer are related by inversion symmetry and hence give the identical $^{51}$V NMR spectrum in the $ac$ plane. As shown in Fig. \ref{Fig6}, we observed two sets of the sharp $^{51}$V NMR spectra, which are related by mirror symmetry. The spectrum exhibits strong angular dependence as the magnetic field is rotated around the principal axes, $X, Y$, and $Z$. Here the axes are nearly identical to those in the $R$ phase and related to the crystal axes ($a_M, b_M, c_M$), as shown in Fig. \ref{Fig1}(b). 

	\begin{figure}
	\includegraphics[scale=0.48]{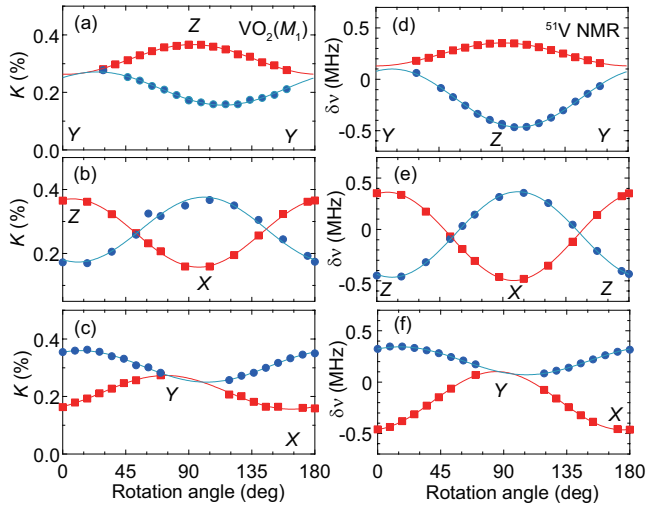}
	\caption{\label{Fig7} 
	Angular dependence of (a-c) $^{51}$V Knight shift $K$ and (d-f) quadrupole splitting frequency $\delta \nu$ in the $M_1$ phase of VO$_2$ at 300 K. Solid curves are fitting results with Eq.(\ref{Eq5}). The red and blue symbols represent the data from two vanadium sites related by mirror symmetry. We could not observe the spectrum at certain directions around the $Y$ axis because the rf-field becomes parallel to the leading magnetic field. 
	}
	\end{figure}
	The narrow linewidth at each angle indicates the vanishing $\chi^{\rm s}$ and spin-echo decay rate $T_2^{-1}$ in the nonmagnetic insulating phase. Furthermore, the maximum quadrupole splitting frequency in the $M_1$ phase doubles compared to the $R$ phase, manifesting an emergence of the large EFG governed by a lower symmetry VO$_6$ distortion. In particular, the displacement of the vanadium atom occurs from the center of VO$_6$ octahedra, which lowers the $d_{||}$ energy level by the ligand field splitting. The quadrupole splitting exhibits a maximum at $[0{\bar 1}1]$ (the $Z$ axis), reaching $|\delta \nu|$ = 0.49(4) MHz. 
	
	In Fig. \ref{Fig7}, we plot the angular dependences of $K$ and $\delta \nu$ in the $M_1$ phase. They exhibit similar behavior for the rotation around the principal axes, indicating the orbital state mostly governed by the crystal field. The data are well fitted with sinusoidal functions of Eq. (\ref{Eq5}) and Eq. (\ref{Eq7}), which directly lead to the Knight shift tensor ${\sf K}$ as 
\begin{equation}
{\sf K} =
%\begin{align}
	\begin{pmatrix}
	    0.37&0&0\\0&0.27&0\\0&0&0.15\\
	\end{pmatrix}
	\% .
	\label{Eq8}
%\end{align}
\end{equation}
In the insulating phase having small $\chi^s$, the Knight shift is dominated by the Van-Vleck orbital susceptibility $\chi^{\rm orb}$. As $\chi^{\rm orb}$ comes from orbital excitations, the ${\sf K}$ anisotropy reflects the $d$ orbital state. The Knight shift tensor gives $K_{\rm iso}^{\rm orb} = 0.26(4)\%$, $K_{\rm ax}^{\rm orb} = -0.11(3)\%$, and $K_{\rm ani}^{\rm orb} = -0.05(2)\%$. We find that $K_{\rm iso}^{\rm orb}$ is also suppressed in comparison with the metallic phase. The result provides the axial anisotropy $K_{\rm ax}^{\rm orb}/K_{\rm iso}^{\rm orb} = -0.42(4)$, which corresponds to the $d_{||}$ orbital polarization of 70\% assuming the hyperfine anisotropy similar to the dipole interaction. 
	
	The angular dependence of $\delta \nu$ in Fig. \ref{Fig7}(b) gives the diagonalized $\delta \nu$ tensor,
    \begin{equation}
\delta \nu =
%\begin{align}
	\begin{pmatrix}
	    373&0&0\\0&115&0\\0&0&-488\\
	\end{pmatrix}
	{\rm kHz}.
	\label{Eq9}
%\end{align}
\end{equation}
The obtained asymmetry $\eta = 0.53$ is smaller than that of the metallic phase, while the maximum of $\delta \nu$ becomes more than twice. We obtained the $d$ orbital contribution to the quadrupole frequency as $\delta \nu^{\rm d} = (393, 405, -798)$ kHz after subtracting $\delta \nu^{\rm ion} = (-2, -29, 31)$ kHz arising from the EFG of surrounding ions using the crystal structure at 300 K \cite{Andersson} and the anti-shielding factor of 10. The experimental result of $\delta \nu$ qualitatively agrees with the EFG obtained from the DFT calculation ${\sf V} = (-0.99, -3.16, 4.14) \times 10^{21}$ V/m$^2$ or $\delta \nu = (273, 85, -357)$ kHz. From Eq. (\ref{Eq2}), the orbital occupation ratio is determined as $d_{||}: d_{yz}: d_{zx} = 0.71(4) : 0.15(3) : 0.14(3)$. The result clearly shows that the $d$ orbital order across the metal-insulator transition. 
	
\section{Discussion}
In this section, based on the result of OR-NMR, we discuss the role of electron correlations, ligand fields, and spin-orbit coupling on the metal-insulator transition for VO$_2$. 

	The $d$ orbital occupation obtained from the present $^{51}$V NMR study for VO$_2$ is summarized in Table \ref{TableI}. 
	We employ three orthogonal $t_{2g}$ bases under the orthorhombic VO$_6$ distortion ($d_\parallel$ or $d_{xy}$, $d_{yz}$, $d_{zx}$) and compare with the other experiment \cite{Haverkort} and theoretical calculations \cite{Laad, Weber, Biermann, Tanaka, Haverkort, Kim, Yuan}. 
	The anisotropy of the magnetic hyperfine coupling or the Knight shift measures the orbital dependent susceptibility, while the EFG anisotopy or the nuclear quadrupole frequency reflects the net $3d$ occupation below the Fermi level. They can sensitively depend on the details of the band structure and spin correlation, as discussed below. In the $R$ phase with the tetragonal or orthorhombic ligand field, we can simply assume the orbital occupation for three $t_{2g}$ orbitals. In the $M_1$ phase, the strong lattice distortion and hybridization between $d$ orbitals and oxygen $p$ orbitals may cause the modification and reconstruction of the wavefunction, as shown in Fig. \ref{Eq8}.

    In the metallic $R$ phase, the Knight shift anisotropy shows the orbital dependent local spin susceptibility: the $d_{yz}$ contribution is largest, while the $d_{\parallel}$ one is smallest. It seems inconsistent with the narrower width of $\pi^*$ bands in the theoretical calculation \cite{Laad, Biermann, Kim, Yuan, Zheng}, although the partial density of states sensitively changes depending on the method of the calculation and the crystal structure. In reference to the result of $\delta \nu^d$, $d_{\parallel}$ can be the most occupied orbital, consistent with the several band calculation \cite{Laad, Biermann, Kim, Yuan, Zheng}. The occupation is naturally expected from the crystal field of the VO$_6$ octahedra elongated along the $Z$ axis in Fig. \ref{Fig1}. The suppressed spin susceptibility of $d_{\parallel}$ may be attributed to orbital-dependent spin correlations such as valence-bond fluctuations along the chain, which leads to the reduced spin entropy and hence the low thermal conductivity \cite{Lee}. 
    
\begin{table}[h]
\begin{tabular}{cccc}
\hline 
Methods &$R$ phase & $M_1$ phase & Ref. \\ 
                 &$d_{||}: d_{yz}: d_{zx}$& $d_{||}: d_{yz}: d_{zx}$ &  
\\ \hline
Knight shift & 0.27 : 0.41: 0.32 & 0.7 : 0.1 : 0.2 & \\ 
hyperfine & 0.23 : 0.44 : 0.33 &  & \\ 
electric quadrupole & 0.49 : 0.24 : 0.26 & 0.71 : 0.15 : 0.14 & \\ 
XAS & 0.33 : 0.51 : 0.16 & 0.81 : 0.09 : 0.10 & \cite{Haverkort} \\ 
\hline
ED & 0.33 : 0.33 : 0.33 & 1 : 0 : 0 & \cite{Tanaka} \\ 
LDA & 0.43 : 0.67 : 0.35 & 0.64 : 0.41 : 0.39 & \cite{Haverkort} \\ 
LSDA $+ U$ & 0.20 : 0.97 : 0.24 & 0.89 : 0.25 : 0.23 & \cite{Haverkort} \\ 
LDA & 0.34 : 0.33 : 0.33 & 0.36 : 0.32 : 0.32 & \cite{Laad} \\ 
DFT+DMFT & 0.37 : 0.25 : 0.33 &    & \cite{Weber} \\ 
LDA & 0.36 : 0.32 : 0.32$^*$ & 0.74 : 0.12 : 0.14 & \cite{Biermann} \\ 
C-DMFT & 0.42 : 0.29 : 0.29 & 0.80$^*$ : 0.10 : 0.10 & \cite{Biermann} \\ 
DFT & 0.46 : 0.32 : 0.22$^*$ &  & \cite{Kim}\\ 
DFT + $U$ & & 0.73 : 0.14 : 0.13$^*$  & \cite{Kim}\\
LDA & & 0.60 : 0.08 : 0.12 & \cite{Yuan}\\
LDA + $U$ & & 0.70 : 0.02 : 0.04 & \cite{Yuan}\\
LDA + $\Delta V$ & & 0.64 : 0.02 : 0.05 & \cite{Yuan}\\
DFT (GGA) & 0.47 : 0.42 : 0.11 & 0.86 : 0.11 : 0.03 & \\
DFT (GGA) + U & & 0.94 : 0.05 : 0.01 & \\
\hline
\end{tabular}
\caption{\label{TableI}Orbital occupation ratio obtained from experiments and theoretical calculations, where ED, L(S)DA, and C-DMFT represent exact diagonalization, DFT $+$ local (spin) density approximation, and cluster dynamical mean field theory, respectively. The DFT (GGA) calculation was performed in the present study. The data marked by $^*$ denotes the partial density of states at the Fermi level. } 
\end{table}

    Entering into the $M_1$ phase, the $d_{||}$ occupation reaches $70\%$ of the full polarization. The result agrees with the XAS measurement \cite{Haverkort}. The band structure calculations also shows predominant $d_{||}$ occupation \cite{Haverkort, Kim, Biermann, Weber, Laad, Yuan}, which is enhanced as the strength of electron correlation increases. The $d_{||}$ orbital order is not necessarily favorable in Mott insulator but facilitates the one-dimensionality linked to the Peierls instability, as discussed in the orbitally-induced or correlation assisted Peierls transition scenario \cite{Haverkort, Biermann}.

%\begin{table}[h]
%\begin{tabular}{ccc}
%\hline 
%phase & $V_X, V_Y, V_Z$ ($10^{21}$ V/m$^2$) & $\eta$ \\ \hline
%$R$ phase ($U=0)$& 0.58, 1.98, $-2.56$ & 0.546\\ 
%$M_1$ phase $(U=0)$& 0.18, 1.78, $-1.96$ & 0.817\\ 
%$M_1$ phase ($U=4.2$ eV) & 0.01, 4.33, $-4.35$ & 0.994\\
%\hline
%\end{tabular}
%\caption{\label{TableII}Electric field gradient  calculated with a DFT method for the $R$ and $M_1$ phases based on the crystal structures at 350 and 300 K, respectively. } 
%\end{table}

Unlike the simple dimerization in spin-Peierls or Peierls transition, the lattice distortion occurs in a complex manner for the $M_1$ phase: the vanadium atom is dislocated perpendicular to the chain from the original position. Such a displacement favors the $d_{||}$ order, as known in the other vanadium oxides such as V$_6$O$_{13}$ \cite{Shimizu}. In the Mott insulating picture, the $d^1$ electron would be occupied in a single orbital by lifting the orbital degeneracy. However, the observed asymmetries in both Knight shift and EFG tensors suggest that the orbital shape is distorted from the simple $d_{||}$ one and originates in an asymmetric wavefunction under the complex ligand field. In this respect, the obtained occupation ratio in Table \ref{TableI} provides the coefficients of the linear combination of the complex wavefunction. 

	\begin{figure}
	\includegraphics[scale=0.3]{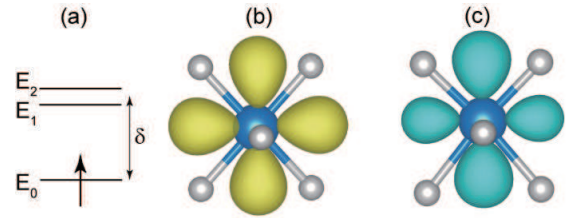}
	\caption{\label{Fig8} 
	(a) Energy diagram of $t_{2g}$ under the nearly tetragonal crystal field for the $R$ and $M_1$ phases. The ground state electron density distribution obtained from the ligand field calculation in (b) the $R$ phase and (c) the $M_1$ phase. EFG tensors have $\eta = 0$ for (b) and $\eta = 0.5$ for (c). 
	}
	\end{figure}

In the ionic limit, the appropriate $d$ electron wavefunction can be evaluated by considering the ligand field in a point-charge approximation. We also take into account an effect of spin-orbit coupling that admixes $d$ orbitals as observed by the Van-Vleck susceptibility. In first order, the ground state wavefunction was obtained by diagonalizing the electronic Hamiltonian including spin-orbit interaction $\lambda${\bf L}$\cdot${\bf S} (the coupling constant $\lambda = 30.7$ meV for V$^{4+}$ \cite{Abragam}) and the nearly tetragonal ligand field splitting $\delta = 76$ meV [Fig. \ref{Fig8}(a)], which are much smaller than the octahedral field $10Dq = 1$ eV. We find that the lowest lying state is governed by the simple $d_\parallel$ orbital for the tetragonal $R$ phase, where $\eta$ of EFG is zero. The population of the $\pi^*$ component increases in the monoclinic $M_1$ phase where the ligand field significantly deviates from the tetragonal symmetry. Correspondingly, the shape of $d$ electron density distribution becomes asymmetric, as manifested in $\eta = 0.5$ [Fig. \ref{Fig8}(c)], consistent with the experimental value. It suggests that the $d$ orbital shape is elongated along the chain direction (the $X$ axis) and shrunk along the direction (the $Y$ axis) normal to the chain. Such an orbital reconstruction under the low-symmetry ligand field can further gain the transfer energy along the chain and stabilize the valence bond formation of the V dimer. 

The above OR-NMR results and analysis provide critical information for the origin of the metal-insulator transition in VO$_2$. We found the orbital occupation change and recombination accompanied by the lattice distortion and charge localization. The observation of the orbital susceptibility implies the residual spin and orbital degrees of freedom in the insulating phase, as expected for Mott insulator. The enhanced spin susceptibility and orbital-dependent correlation are also consistent with the strongly correlated regime. The observed electron cloud distribution supports the $d$ orbital order governed by the local ligand field and the dimerization, compatible to a localized molecular orbital picture modified from the original atomic orbital. Thus the insulating $M_1$ phase has a character of the Mott insulator with strong dimerization. The observed $d$ orbital occupation switching would trigger the metal-insulator transition, because the system acquires a large spin excitation gap. This picture is compatible to the orbitally driven Mott transition accompanied by the spin-Peierls transition. We stress that the electronic structure for transition-metal compounds with orbital degrees of freedom must be studied beyond the simplified ligand field such as tetragonal and trigonal field. Actually, the symmetry of the ligand field is often complex, and the ground state wavefunction should be expressed as a linear combination of the orthogonal bases. The physics of metal-insulator transition will be reconstructed using the accurate form of the single-electron or multi-electron wavefunction beyond the atomic orbital.

\section{Conclusion}
	$^{51}$V NMR measurements revealed the orbital susceptibility and the $3d$ orbital occupation in vanadium dioxide. In the metallic phase, we observed the significant orbital degeneracy in the $t_{2g}$ manifolds. The obtained ratio for the three-orbital contribution is consistent in the Knight shift and the hyperfine coupling, which measure the partial density of states and orbital-dependent spin correlation. The analysis of the hyperfine tensors and theoretical calculations in the insulating phase suggest that the metal-insulator transition involves the $d$ occupation switching and the reconstruction of the wavefunction, which can be optimized to acquire the spin gap and stabilize the insulating phase. The residual orbital susceptibility implies the spin-orbital degrees of freedom supporting the Mott insulating picture. In contrast to the simple orbital order in Mott insulator, the complex lattice distortion induces the asymmetric form of the localized molecular orbital governed by the ligand field. 

\section*{Acknowledgments}
This work was supported by JSPS KAKENHI (Grants No. JP19H01837, JP16H04012, and JP19H05824).

\bibliography{VO2_bib}
	
\end{document}